\documentclass{aastex63}

\usepackage{amsmath}

\received{2020 September 9}
\revised{2020 December 3}
\accepted{2020 December 13}
\published{2021 MM DD}

\submitjournal{ApJ}

\begin{document}

\title{Multimessenger parameter estimation of GW170817: from jet structure to the Hubble constant}

\correspondingauthor{Hao Wang}
\email{wang4145@purdue.edu}

\author{Hao Wang}
\affiliation{Department of Physics and Astronomy, Purdue University, 525 Northwestern Avenue, West Lafayette, IN 47907, USA}

\author{Dimitrios Giannios}
\affiliation{Department of Physics and Astronomy, Purdue University, 525 Northwestern Avenue, West Lafayette, IN 47907, USA}

\begin{abstract}

The electromagnetic radiation that followed the neutron star merger event GW170817 revealed that gamma-ray burst afterglows from jets misaligned with our line of sight exhibit a light curve with slowly rising flux.  The slope of the rising light curve depends sensitively on the angle of the observer with respect to the jet axis, which is likely to be perpendicular to the merger plane of the neutron star binary. Therefore, the afterglow emission can be used to constrain the inclination of the merging system. Here, we calculate the gamma-ray burst afterglow emission based on the realistic jet structure derived from general-relativistic magnetohydrodynamical simulations of a black hole torus system for the central engine of the gamma-ray burst. Combined with gravitational wave parameter estimation, we fit the multi-epoch afterglow emission of GW170817. We show that with such a jet model, the observing angle can be tightly constrained by multi messenger observations. The best fit observing angle of GW170817 is $\theta_{\rm v} = 0.38\pm 0.02$ rad. With such a constraint, we can break the degeneracy between inclination angle and luminosity distance in gravitational wave parameter estimation, and substantially increase the precision with which the Hubble constant is constrained by the standard siren method. Our estimation of the distance is $D_{\rm L}=43.4\pm 1\ \rm Mpc$ and the Hubble constant constraint is $69.5\pm 4\ \mathrm{km\ s^{-1}\ Mpc^{-1}}$. As a result, multimessenger observations of short-duration gamma-ray bursts, combined with a good theoretical understanding of the jet structure, can be powerful probes of cosmological parameters. 

\end{abstract}


\section{Introduction} \label{sec:Introduction}

Since the detection of the neutron star merger event in gravitational waves (GW170817 \cite{gw170817}), gamma-rays (GRB 170817A \cite{grb170817a}) and optical (AT2017gfo \cite{at2017gfo}), short-duration gamma-ray bursts became one of the most successful targets of multimessenger astrophysics. Combining information from both gravitational waves and electromagnetic waves, an impressive breadth of Astrophysical questions can be probed, such as, for example, the origin of the heavy elements \citep{2017Natur.551...80K}, the neutron star equation of state \citep{2018PhRvL.121p1101A}, testing the weak equivalence principle \citep{2017ApJ...851L..18W}, or the value of standard cosmological parameters \citep{2017Natur.551...85A}. Recently, the Hubble constant has attracted much attention because of the tension between the value measured based on the cosmic microwave background and the cosmic distance ladder methods \citep{planck, SHOES2019, Verde2019}. An alternative method to constrain the Hubble constant, the so-called standard siren method, takes advantage of the luminosity distance measured by gravitational wave signal and the cosmological redshift of the source's host galaxy \citep{2010ApJ...725..496N}. Such a method is independent from the previous methods, and may serve as an additional probe of the expansion rate of the universe. 

The standard siren method requires an astrophysical source that provides bright electromagnetic counterparts along with a gravitational wave detection. One of the most promising sources is a binary neutron star merger which, not only gives a powerful gravitational wave signal, but is also expected to launch relativistic and non-relativistic outflows. The outflows can power a gamma-ray burst, afterglow emission, and a kilonova \citep{Metzger2019}. In this case, gravitational (electromagnetic) waves can be used to derive the luminosity distance (source location), respectively.  However, the luminosity distance of a binary neutron star merger, as determined by the gravitational wave signal, can be strongly degenerate with the orbital inclination of the system; that is to say from the gravitational signal alone, one is hard to distinguish whether a source is further away with the binary orbit facing Earth, or closer but the binary orbit has been highly inclined to the line of sight. Such degeneracy results in rather large uncertainty of distance measurement, making it harder to achieve a precision level comparable to other methods. 

One way of improving the accuracy of the method is by breaking the distance-inclination degeneracy. This can be achieved by including information from the gamma-ray burst afterglow observations. Several studies suggest that the afterglow emission that followed GRB 170817A originates from the interactions of a structured jet (i.e., a jet with its properties like energy and Lorentz factor smoothly varying as a function of polar angle) with the ambient gas\citep{2017MNRAS.472.4953L, 2017Natur.551...71T, 2018NatAs...2..751L, 2018MNRAS.478L..18T, 2018MNRAS.478.4128G, 2018A&A...613L...1D, vla3, 2018ApJ...868L..11M, 2018PhRvL.120x1103L, 2018MNRAS.478..733L, 2018ApJ...863...58X, 2019ApJ...870L..15L, 2019Sci...363..968G, 2019ApJ...883...15G, 2019MNRAS.482.5430B, 2020MNRAS.495.3780N}. In such a scenario, the slope of the rising flux, observed for $\sim$6 months after the burst, depends sensitively on viewing angle of the observer. Assuming that the jet emerges perpendicular to the orbital plane of the merging binary (i.e., the binary inclination angle is the same as the jet observing angle), one can therefore infer the inclination angle by fitting the afterglow data, provided that we have a realistic model for the jet structure. A much better constrained inclination angle by this method will help break the degeneracy and greatly reduce the uncertainty of distance.

Much of the work related to predicting the shape of the gamma-ray burst afterglow as a function of the jet inclination relies on a prescribed model for the jet structure (i.e., Gaussian or power Law dependence for the jet power as a function of polar angle, such as \cite{2019NatAs...3..940H, 2018MNRAS.478L..18T, 2019MNRAS.489.1919T, 2020MNRAS.493.3521B}). These approaches allow for important inferences on how the jet distributes its energy from the afterglow light curve. Not surprisingly, however, the observing angle fitted for in these studies sensitively depends on the assumptions on the jet profile. In a different approach, jets interactions are simulated numerically. The jet is not launched consistently (the central engine is not resolved), but its hydrodynamical interactions with an assumed ambient gas density profile are followed on a large scale where the jet structure becomes independent of distance\citep{2019MNRAS.488.2405G, 2020MNRAS.495.3780N, 2019arXiv190908627M}. This approach, while clearly more self-consistent, still suffers from ambiguities on how the jet is injected and/or the ambient gas is set up in the first place. Clearly, a reliable determination of the jet structure requires a consistent model for the jet launching, the properties of the surrounding gas, as well as the jet interactions with it.

In this work we use the jet structure derived from three-dimensional general-relativistic magnetohydrodynamical (GRMHD) simulation of \cite{2019MNRAS.482.3373F}. This simulation assumes a black hole torus model for the engine of the gamma-ray burst; it follows consistently to accretion into the black hole and the jet formation as a result of the Blanford-Znajek process \citep{BZ77} while outflows from the torus provide the natural envelope of slowly expanding gas that collimates the jet. The jet-wind interaction is followed out a sufficiently large distance for the jet structure (energy and Lorentz factor as functions of polar angle) to be determined \citep{adithan2}. Such a calculation is a step toward physically more reliable model which can reduce the uncertainty of jet opening angle and overall profile. In this work, we use this simulated jet profile to calculate the predicted afterglow light curve and fit the observed afterglow emission \citep{vla1, vla2, vla3, vla4} simultaneously with the gravitational wave observations from GW170817.  We show that in our model for the jet structure, the observing angle can be tightly constrained. Combining with gravitational wave data, the distance uncertainty is also greatly reduced. 

This paper is organized as follows. In \S \ref{sec:Afterglow Model} we introduced our structured jet model and the afterglow calculation details. In \S \ref{sec:fitting} we describe the details of radio data fitting, and how we take advantage of gravitational wave data to perform joint fitting. We show our fitting result in \S \ref{sec:results}, including the implications of some parameters and how observing angle and luminosity distance can be better constrained. In \S \ref{sec:prospect} we briefly discuss the prospect of future multimessenger events. We summarize our conclusion in \S \ref{sec:conclusion}.

\section{Afterglow Model}
\label{sec:Afterglow Model}

\subsection{Jet Structure}
\label{subsec:Jet Structure}

\begin{figure}
	\includegraphics[width=\columnwidth]{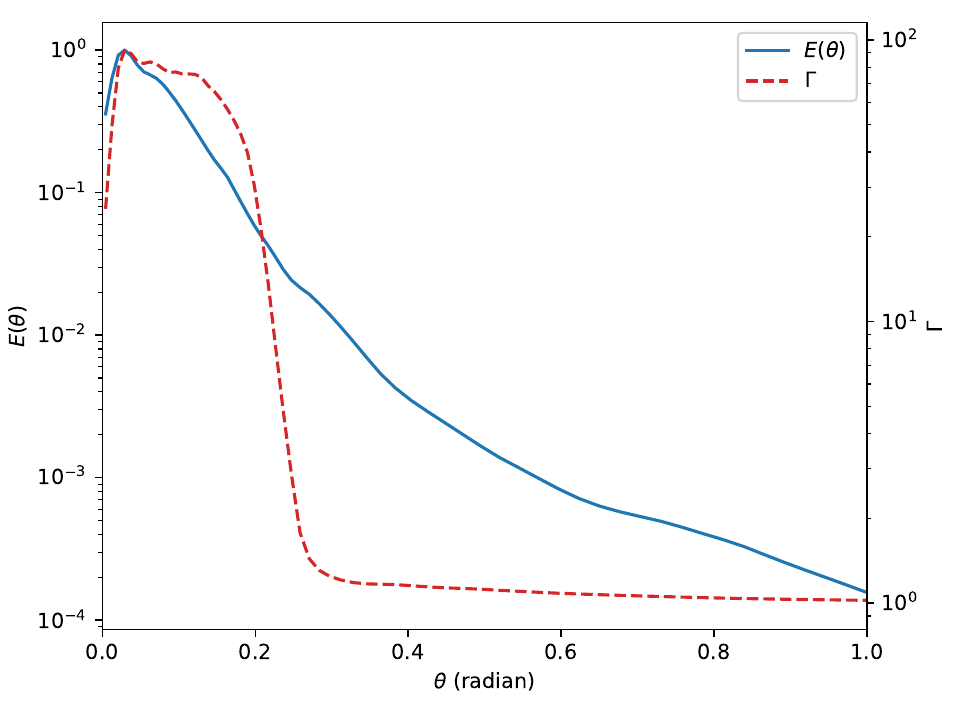}
    \caption{The energy and Lorentz factor as a function of polar angle in our jet model. Energy is normalized to its peak value.}
    \label{fig:jet}
\end{figure}

The afterglow emission is powered by the shocks that the jet drives into the ambient gas. The afterglow light curve strongly depends on the jet profile at large distance from the source where these interactions become substantial. On the other hand, the jet profile is determined at much more compact scale by the interactions of the jet while it breaks out from slower gas expelled within a fraction of a second before and after the merger time.  In this work we adopt a jet structure derived from 3D GRMHD simulation where the jet is launched from a 3 solar mass central black hole with 0.8 spin surrounded by a compact torus of 0.03 solar mass embedded with a poloidal magnetic field. A detailed description of the setup can be seen from \cite{2019MNRAS.482.3373F, adithan2}. At the initial stages of the simulation, the torus is characterized by slower wind-type outflows while mass, and magnetic flux, start accreting into the black hole. Once sufficient magnetic flux has accumulated through the black hole ergosphere, a powerful, relativistic jet forms. The jet is surrounded by the wind from the accretion disk; it interacts with it and colimates out to distances of $\sim$a few 1000 gravitational radii. After the jet breaks out, the jet turns conical and its structure is almost frozen as a function of distance from the central engine. The simulation predicts a tightly collimated relativistic jet with an opening angle of $\sim 0.2$. The angular distribution of initial Lorentz factor and energy of the jet predicted by that work is shown in Figure \ref{fig:jet}. In the following section, we describe how one can calculate the afterglow emission given the jet profile.

\subsection{Dynamical Evolution of the External Shock}
\label{subsec:Shock Dynamical Evolution}

The relativistically moving jet drives a blastwave into the ambient gas. Here, we work in the thin shell approximation, where we assume that the swept-up ambient matter is compressed in a thin region whose scale is much smaller than the radial scale. While the blastwave is initially ultra-relativistic, it gradually slows down and turns trans-relativistic within months after the burst. For a misaligned jet, with its core located at an angle $\theta_{\rm v}>1/\Gamma$ with the line of sight, the emission from the core is initially beamed away from our line of sight. The peak of the afterglow emission takes place when the Lorentz factor of the core of the blast slows down to $\Gamma\sim 1/\theta_{\rm v}$ which falls into mildly relativistic domain if $\theta_{\rm v}$ is relatively large. Since we wish to follow the blast emission up to, and shortly after, its peak, a treatment that follows the blast deceleration down to a trans-relativistic speed is required. In this work we use a semi-analytical model following the shock jump conditions derived by an approximate trans-relativistic EoS \citep{Uhm2011}. This model works well at both high and modest Lorentz factors and can smoothly transform to a Sedov-Taylor solution when the shock is no longer relativistic.

Our model assumes one-dimensional radial solution for each blast segment. Therefore, we neglect sideways (lateral) expansion in our model. The sideways expansion of the blast is important only at later stages of the blast evolution and it affects the light curve predictions after its peak \citep{2018MNRAS.478.4128G}. However, the key factor to constrain inclination angle, of interest in this study, is the arising slope before the peak flux, where the jet is still relativistic. So the sideways expansion will not influence our result.

To describe the physical quantities more clearly, here we consider two reference frames. One is burst rest frame (hereafter BR frame) which is at rest with the central engine. The other is the shock co-moving frame (hereafter CM frame) which is at rest with respect to the gas just downstream of the forward shock. Primed variables (i.e., denoted with $'$) are measured in CM frame, while non-primed ones are measured in BR frame. 

Under the approximation of neglecting sideways expansion, we can regard the jet evolution at each radial direction as being independent. It is therefore convenient to describe the evolution at each direction by using isotropic equivalence quantities. Consider an isotropic burst of total energy $E_{\rm iso} = \Gamma_0 M_{\rm ej}c^2$, where $M_{\rm ej}$ is the mass of ejecta and $\Gamma_0$ is its initial Lorentz factor. The outflow spreads in a cold uniform environment of number density $n$ and forms a relativistic shock. We assume that the blastwave is expanding adiabatically without further energy injection (for considerations of energy injection in the blast see \cite{2020ApJ...899..105L}). Energy loss due to radiation is negligible compared with $E_{\rm iso}$. We define the radius $R$ of the shock by the distance from burst center to the forward shock. All swept-up mass is compressed in the downstream of forward shock, moving with Lorentz factor $\Gamma$ measured in BR frame, and its isotropic kinetic energy is $E_{\rm k}$. We also assume that all ejecta are moving together with shocked ambient at the same Lorentz factor behind the contact discontinuity, and its total energy is $E_{\rm ej} = \Gamma M_{\rm ej}c^2$.

The trans-relativistic shock jump conditions is \citep{Uhm2011, Ryan2019}

\begin{eqnarray}
n' & = & 4 \Gamma n \label{eq:shock1} \\
e'_{\rm th} & = & 4\Gamma (\Gamma - 1)nm_p c^2 \label{eq:shock2}\\
p' & = & 4(\Gamma^2-1)nm_p c^2/3 \label{eq:shock3}\\
\dot{R} & = & 4\beta c \Gamma^2/(4\Gamma^2-1), \label{eq:radius}
\end{eqnarray}

where $n'$, $e'_{\rm th}$ and $p'$ are the number density, internal energy, and pressure of shocked region, respectively. $\beta = (1-\Gamma^{-2})^{1/2}$ is the dimensionless speed. $c$ is the speed of light. Because all swept-up matter is gathered in the shell, one can solve the width of the shell by equating $4\pi R^3n/3=4\pi R^2\Delta R(n'\Gamma)$. Applying Eq. \ref{eq:shock1} we have
\begin{equation}
	\Delta R = R/(12\Gamma^2).
\end{equation}

The kinetic energy is therefore \citep{Uhm2011, Ryan2019, 2013arXiv1309.3869V}
\begin{eqnarray}
	E_{k} & = & \left[ (\rho' c^2 + e'_{th} + p')\Gamma^2 - p' - \Gamma \rho' c^2\right]4\pi R^2\Delta R \nonumber \\
	& = & 4\pi R^3 n m_p \beta^2(4\Gamma^2 - 1)c^2/9,
\end{eqnarray}

where $\rho'=n'm_p$ is the density of shocked region.

As the shock spreads, the ejecta gives its kinetic energy to swept-up matter while keeping the total energy conserved. i.e. $E_{\rm iso} = E_{\rm k} + E_{\rm ej} = \mathrm{const}$. Substituting $M_{\rm ej}$ in terms of $E_{\rm iso}$ and $\Gamma_0$, the total energy is
\begin{equation}
   \label{eq:energy}
	E_{\rm iso} = \frac{1}{1 - \Gamma/\Gamma_0}\frac{4\pi}{9}R^3 nm_p\beta^2(4\Gamma^2 - 1)c^2. 
\end{equation}

The shock dynamical evolution can then be solved by combining Eq \ref{eq:radius} and Equation \ref{eq:energy}. The solution shows three parts of the evolution. Initially the coasting phase where most of the energy is carried by the ejecta and its Lorentz factor remains almost constant. When the accumulated ambient mass is sufficiently large, the shock enters the deceleration phase, where $\Gamma \propto R^{-3/2}$, while $R \sim ct$ because the shock is still ultra-relativistic. When the shock is no longer relativistic, we reproduce that $R \propto t^{2/5}$ as expected from the Sedov-Taylor solution.

\subsection{Synchrotron Radiation}
The accelerated, non-thermal electrons in the shock produce synchrotron radiation. In the standard GRB afterglow model, the non-thermal electron Lorentz factor distribution can be characterized by a power-law function $n'_e = Cn' \gamma_e^{-p},\ \gamma_e>\gamma_m$, where $\gamma_m$ is the minimum electron Lorentz factor and $C$ is a normalization factor. $C$ and $\gamma_m$ can be solved assuming that a fraction $\epsilon_e$ of internal energy is transferred to non-thermal electrons \citep{Sari1998}.
In this work we also use an additional free parameter $f$ which describes the fraction of accelerated electrons to the total number in the swept-up region, i.e. $n'_e = C fn'\gamma_e^{-p}$ \citep{2005ApJ...627..861E, 2013arXiv1309.3869V}.

With such modification, the electron minimum Lorentz factor is
\begin{equation}
    \label{eq:gamma_m}
	\gamma_m = \frac{p-2}{p-1}\frac{\epsilon_e m_p}{fm_e}(\Gamma-1)
\end{equation}

Assume a fraction $\epsilon_B$ part of internal energy transferred to that of the magnetic field, the magnetic field therefore is
\begin{eqnarray}
	B' & = & \sqrt{8\pi e'_{th}\epsilon_B} \nonumber \\
	& = & \sqrt{32\pi\Gamma(\Gamma-1)nm_p\epsilon_B c^2}.
\end{eqnarray}

The cooling Lorentz factor of the electrons is found by equating the synchrotron loss to the expansion time of the blast, resulting in
\begin{equation}
	\gamma_c = \frac{6\pi m_e\Gamma c}{\sigma_T B'^2 t},
\end{equation}
where $t$ is the time since the burst measured in BR frame.

For each characteristic Lorentz factor $\gamma_i=\gamma_m,\ \gamma_c$, the corresponding frequency is
\begin{equation}
	\nu'_i = \frac{3eB'\gamma_i^2}{4\pi m_e c}.
\end{equation}

The synchrotron radiation peak emissivity is
\begin{equation}
	\epsilon'_P = \frac{\sqrt{3}e^3 B' f n'}{m_e c^2}.
\end{equation}

The broad band observation of GRB 170817A from radio to X-ray shows a constant spectral index \citep{vla2, vla3}. This spectrum is consistent with optically thin synchrotron radiation.\footnote{For consistency, we have included the synchrotron self-absorption (SSA) in our calculation, but the SSA break is below the observed band, so for simplicity, we do not include SSA here.}. The Spectrum is then
\begin{equation}
	\epsilon'_{\nu'} = \epsilon'_P\times
	\begin{cases}
		(\nu' / \nu'_m)^{1/3} & \nu' < \nu'_m < \nu'_c \\
		(\nu' / \nu'_m)^{-(p-1)/2} & \nu'_m < \nu' < \nu'_c \\
		(\nu'_c / \nu'_m)^{-(p-1)/2} (\nu' / \nu'_c)^{-p/2} & \nu'_m < \nu'_c < \nu' \\
		(\nu' / \nu'_m)^{1/3} & \nu' < \nu'_c < \nu'_m \\
		(\nu'_c / \nu'_m)^{1/3} (\nu' / \nu'_c)^{-1/2} & \nu'_c < \nu' < \nu'_m \\
		(\nu'_c / \nu'_m)^{1/3} (\nu'_m / \nu'_c)^{-1/2} (\nu' / \nu'_m)^{-p/2} & \nu'_c < \nu'_m < \nu'.
	\end{cases}
\end{equation}

\subsection{Observed Flux of a structured jet}
In order to calculate the total observed flux, we need to integrate over the whole solid angle. It is convenient to consider a spherical coordinate system where the origin is the center of the burst and z-axis is the axis of the jet. It is also natural to set the observer's direction as $\varphi = 0$. In the structured jet case, Lorentz factors, radius, and emissivity are functions of $\theta$ and $t$ and can be solved given the initial condition at angle $\theta$. The isotropic equivalent energy at each direction can be calculated by a free parameter $E_{\rm tot}$ which is the angle integrated jet total energy. i.e. $E_{\rm iso}(\theta) = 4\pi E(\theta)E_{\rm tot}/\int{E(\theta)d\theta}$. The isotropic equivalent radiation power at coordinate $(\theta, \varphi)$ is then
\begin{equation}
    P_{\rm iso}(\theta, \varphi, t) = \epsilon'_{\nu'}(\theta, t)4\pi R(\theta, t)^2\Delta R'(\theta, t)\delta(\theta, \varphi, t)^3,
\end{equation}
where $\Delta R'=\Delta R\Gamma$ is the shock width measured in CM frame, $\delta = [\Gamma(\theta, t)(1-\beta(\theta, t)\mu(\theta, \varphi)]^{-1}$ is Doppler factor, $\mu$ is the cosine angle between $(\theta, \varphi)$ and the observer's direction 
\begin{equation}
    \mu = \cos\theta\cos\theta_{v} +\sin\theta\cos\varphi\sin\theta_{v}.
\end{equation}
The observed flux at frequency $\nu_{\rm obs}$ and observer's time $T_{\rm obs}$ since burst is therefore
\begin{equation}
	F_{\nu_{\rm obs}}(T_{\rm obs}) = \frac{1+z}{4\pi D_L^2}\int \frac{P_{\rm iso}(\theta, \varphi, t)}{4\pi}\mathrm{d}\Omega,
\end{equation}
where $z$ is redshift and $D_L$ is the luminosity distance.

The frequency at CM frame can be calculated through Doppler factor $\nu' = \nu_{\rm obs}(1+z)/\delta$. The time $t$ at BR frame is numerically solved by equal-arrival-time-surface provided $\theta$, $\varphi$ and $T_{\rm obs}$
\begin{equation}
	t - R(\theta, t)\mu(\theta, \varphi)/c = T_{\rm obs}/(1+z).
\end{equation}

\subsection{Light Curve Behavior}
\begin{figure}
	\includegraphics[width=\columnwidth]{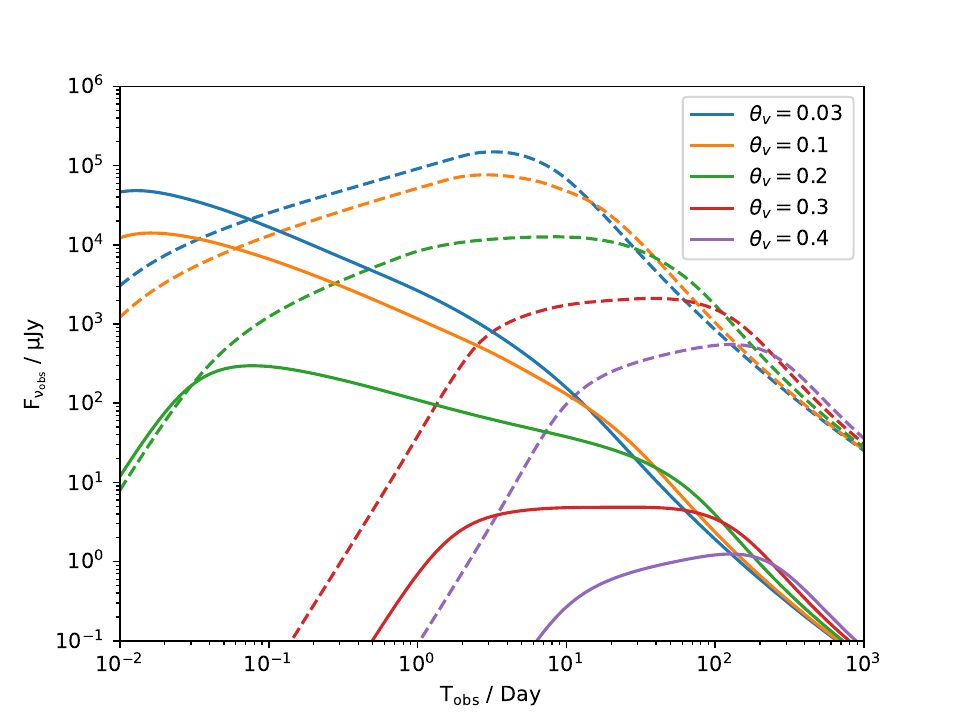}
    \caption{The light curves at different observing angles generated by our model. The parameter set is $E_{\rm tot} = 10^{50}\mathrm{erg}$, $n = 10^{-3}\mathrm{cm^{-3}}$, $\epsilon_{\rm e} = 0.1$, $\epsilon_{\rm B} = 10^{-3}$, $D_L = 40 \mathrm{Mpc}$, $f = 1$. The solid (dashed) lines show the  $10^{14}\mathrm{Hz}$ ($3\mathrm{GHz}$) light curves, respectively. Angles are measured in rad.}
    \label{fig:lc}
\end{figure}

In Fig. \ref{fig:lc}, we plot a collection of light curves of the afterglow emission predicted by our model for a jet observed at different angles. At $10^{14}$Hz (solid lines), the light curves are characterized by two breaks. The early-time break represents the transition from coasting stage to deceleration stage of the segment of the jet that dominates the early-time emission. After that stage, because of the beaming effect as the jet slows down, observers can see a gradually increasing area closer to the core of the jet. The slope of the light curve at this stage strongly depends on $\theta_v$, and is a key quantity that we constrain with the fits it in our work. When the observer sees the emission from the jet core, the most powerful segment of the jet contributes, and dominates, the observed flux and the light curves hereafter have a second break. 

At 3GHz, the light curves of low $\theta_{\rm v}$ at early time show a feature of increasing flux, which is different from ones from the same angle but higher frequency. This feature originates from ``spectral break,", i.e., the source frequency $\nu' = \nu_{\rm obs}(1 + z)/\delta$ is lower than $\nu'_m$ because of the high Doppler factor.

\section{Parameter Estimation}
\label{sec:fitting}
In this section, we apply our model to the radio data of GRB 170817A and perform a Bayesian parameter estimation. The data points that we use for our fits are gathered from a number of publications \citep{vla1, vla2, vla3, vla4}. Higher energy data like in the optical and X-ray bands follow the power-law extrapolation of the radio data \citep{vla3}. As such, they can help to constrain electron power-index $p$, but do not provide additional information to constrain $\theta_{\rm v}$ and are unnecessary in our work. We do, however, check for self consistency that our best fit models also account for the X-ray observations (e.g., the cooling frequency is not crossing the X-ray band during the observational window; see \S \ref{sec:results} for more details). To take advantage of the multi-wavelength information, we fix $p=2.17$ in our fitting. We also fix the source cosmological redshift to $z = 0.0098$. The free parameters to calculate the flux at a given frequency $\nu_{\rm obs}$ in our afterglow model are then ($E_{\rm tot}$, $\theta_{\rm v}$, $n$, $\epsilon_{\rm e}$, $\epsilon_B$, $D_L$, $f$). We calculate the fitting goodness by $\chi^2$ and the likelihood function is then $\mathcal{L} = \exp(-\chi^2/2)$. It is generally not possible to constrain the distance solely by afterglow data without cosmological assumptions. However, combining with gravitational waves, we can not only include distance information but also break the degeneracy between distance and inclination angle. Such constraint is independent of cosmological parameters. 

To take advantage of gravitational waves information, we use the open package Bilby \citep{2019ApJS..241...27A} to perform a gravitational wave parameter estimation of GW170817 in advance. The gavitational wave data of GW170817 are collected from GW Open Science Center\footnote{https://www.gw-openscience.org/eventapi/html/O1\_O2-Preliminary/GW170817/v2/}. We have used the cleaned version where the glitch has been removed. The algorithm is described in \cite{2019ApJS..241...27A} and the priors are chosen to be the same as \cite{2019PhRvX...9a1001A}, but with sky localization fixed to its host galaxy NGC4993. A fixed coordinate will help reduce luminosity distance uncertainty. The waveform template we used here is IMRPhenomPv2\_NRTidal \citep{2017PhRvD..96l1501D,2019PhRvD..99b4029D,2014PhRvL.113o1101H}. The posterior distribution of luminosity distance and inclination angle then serve as a prior to the corresponding parameters in our model, where we have assumed that the inclination angle is exactly the observing angle (i.e., that the jet is ejected perpendicular to the plane of the merger). To deal with the facing-off cases (i.e., the inclination angle is larger than $\pi/2$ in the gravitational wave analysis), we adjust them to $\pi - \theta_{\rm v}$ in the afterglow calculation.

To show how model parameters work in the radio data fittings and how we combine our model with gravitational waves, we did three fitting with different prior sets. They are summarized in Table \ref{tab:prior}. Below are the description of those priors.

\begin{table}
	\centering
	\caption{The prior distribution in parameter Estimation.}
	\label{tab:prior}
	\begin{tabular}{cccc} % four columns, alignment for each
		\hline
		\hline
		Sampling Parameters & Prior Set 1 & Prior Set 2 & Prior Set 3 \\
		\hline
		$\log(E_{\rm tot}/\mathrm{erg})$ & [47, 55] & [47, 55] & [47, 55] \\
		$\cos{\theta_{\rm v}}$ & [0, 1] & [0, 1] & GW \\
		$\log(n/\mathrm{cm^{-3}})$ & [-6, 0] & [-6, 0] & [-6, 0] \\
		$\log(\epsilon_{\rm e})$ & [-3, 0] & [-3, 0] & [-3, 0] \\
		$\log(\epsilon_{\rm B})$ & [-6, 0] & [-6, 0] & [-6, 0] \\
		$D_L/\mathrm{Mpc}$ & 40 & 40 & GW \\
		$\log(f)$ & 0 & [-4, 0] & [-4, 0] \\
		\hline
	\end{tabular}
	\tablecomments{All the distributions are uniform in the given bound. GW refers to the kernel density estimation built from posterior samples of gravitational waves.}
\end{table}

\begin{description}
\item[\it Prior set 1.] This is the simplest afterglow fit setting where the fraction of accelerated electrons $f$ is set to 1 and distance $D_L$ is fixed to 40 Mpc. Other parameters are uniformly distributed in bounds.
\item[\it Prior set 2.] In this prior set we leave $f$ free to show the importance of this parameter in our fit, while the distance is still fixed.
\item[\it Prior set 3.] We leave all parameters free, while the prior of observing angle and luminosity distance is a kernel density estimation function built from the posterior samples of GW170817 in our analysis above. Applying this prior is equivalent to a joint parameter estimation of GRB 170817A and GW170817.
\end{description}

\begin{figure}
	\includegraphics[width=\columnwidth]{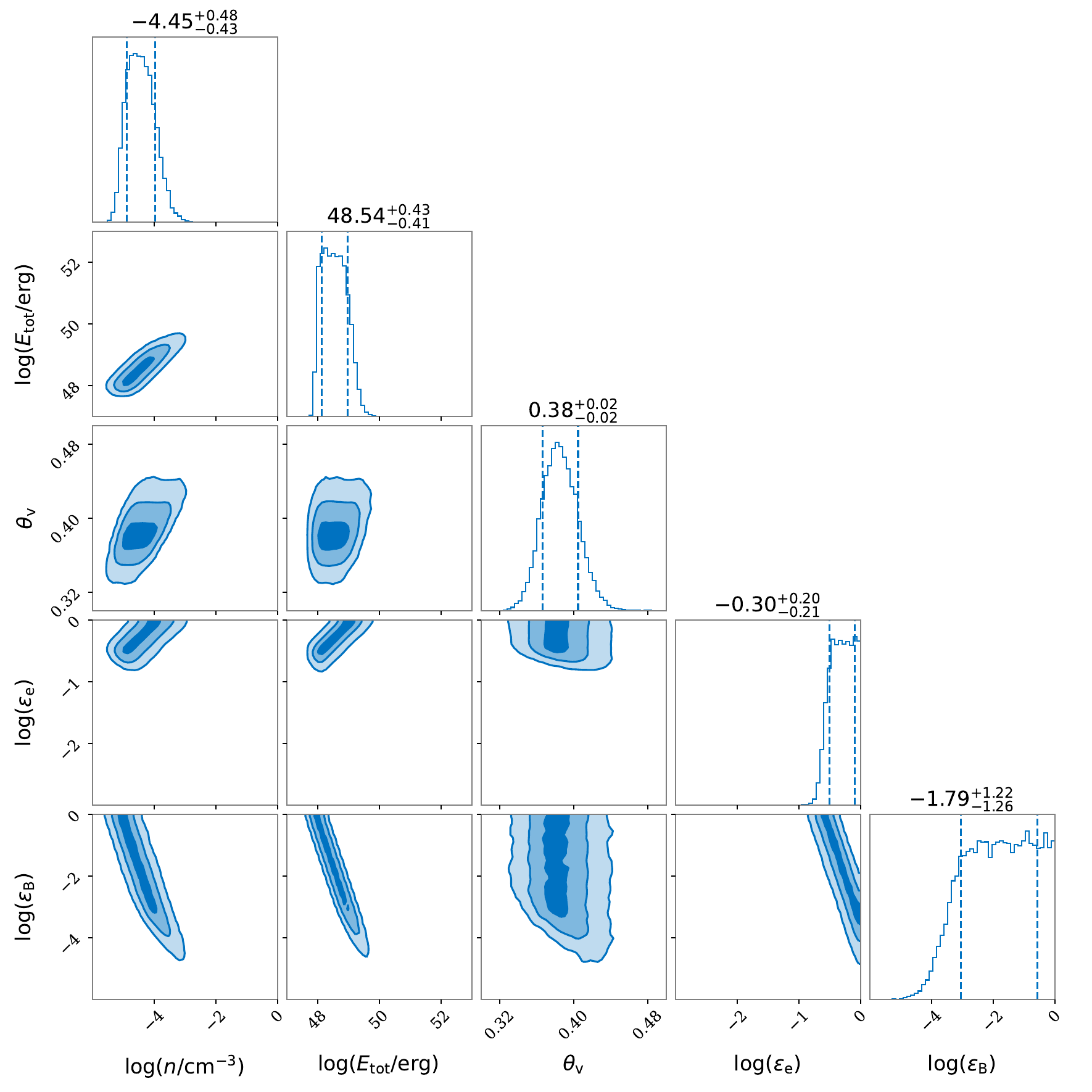}
    \caption{Parameter estimation result with prior set 1, where $f=1$ and $D_{\rm L}=40\ \rm Mpc$. The contours with different colors represent $1\sigma-3\sigma$ uncertainty levels.}
    \label{fig:pe_fixed_f}
\end{figure}

\begin{figure}
	\includegraphics[width=\columnwidth]{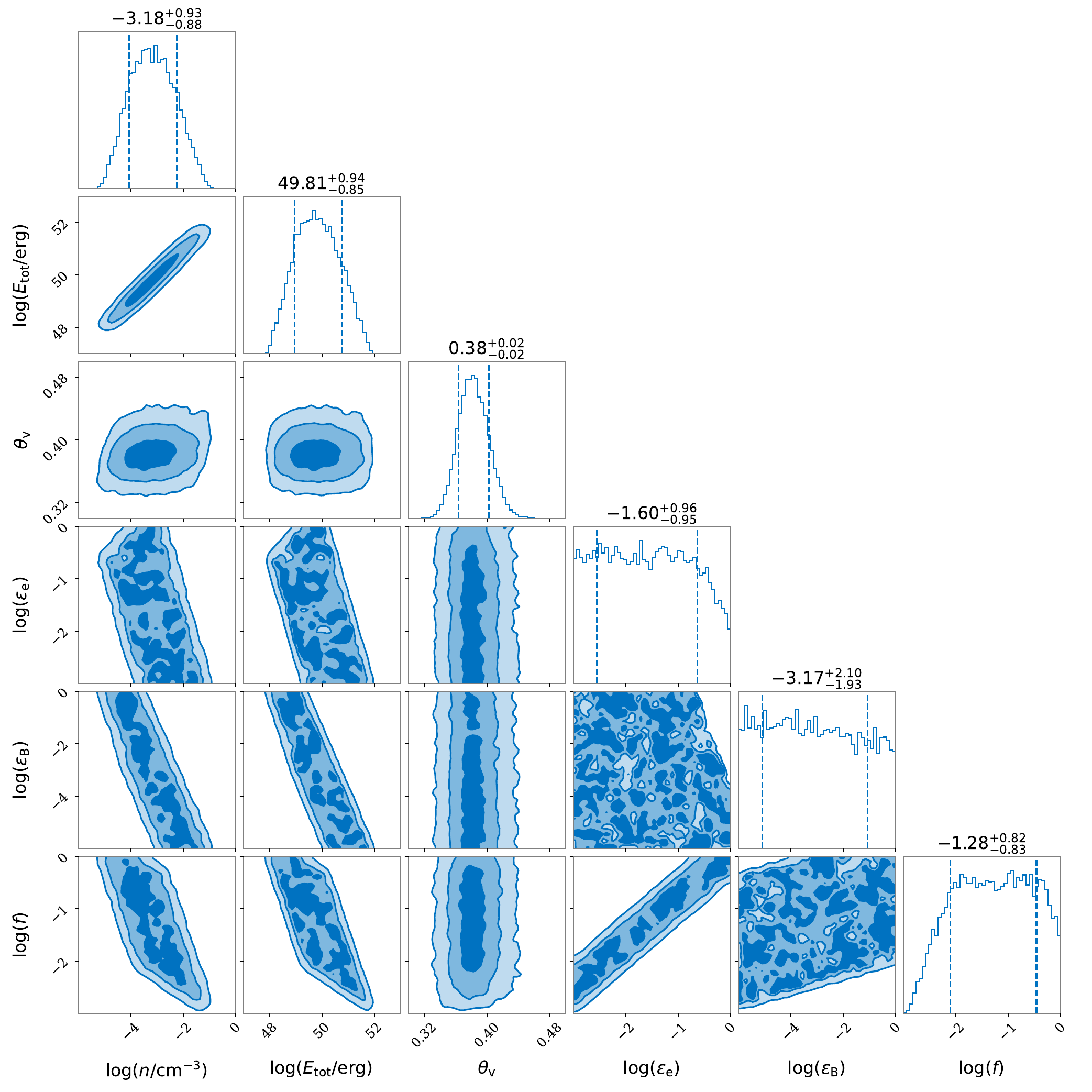}
    \caption{Parameter estimation result with prior set 2, where $f$ is left free while $D_{\rm L}$ is still fixed to 40 Mpc.}
    \label{fig:pe_free_f}
\end{figure}

\begin{figure}
	\includegraphics[width=\columnwidth]{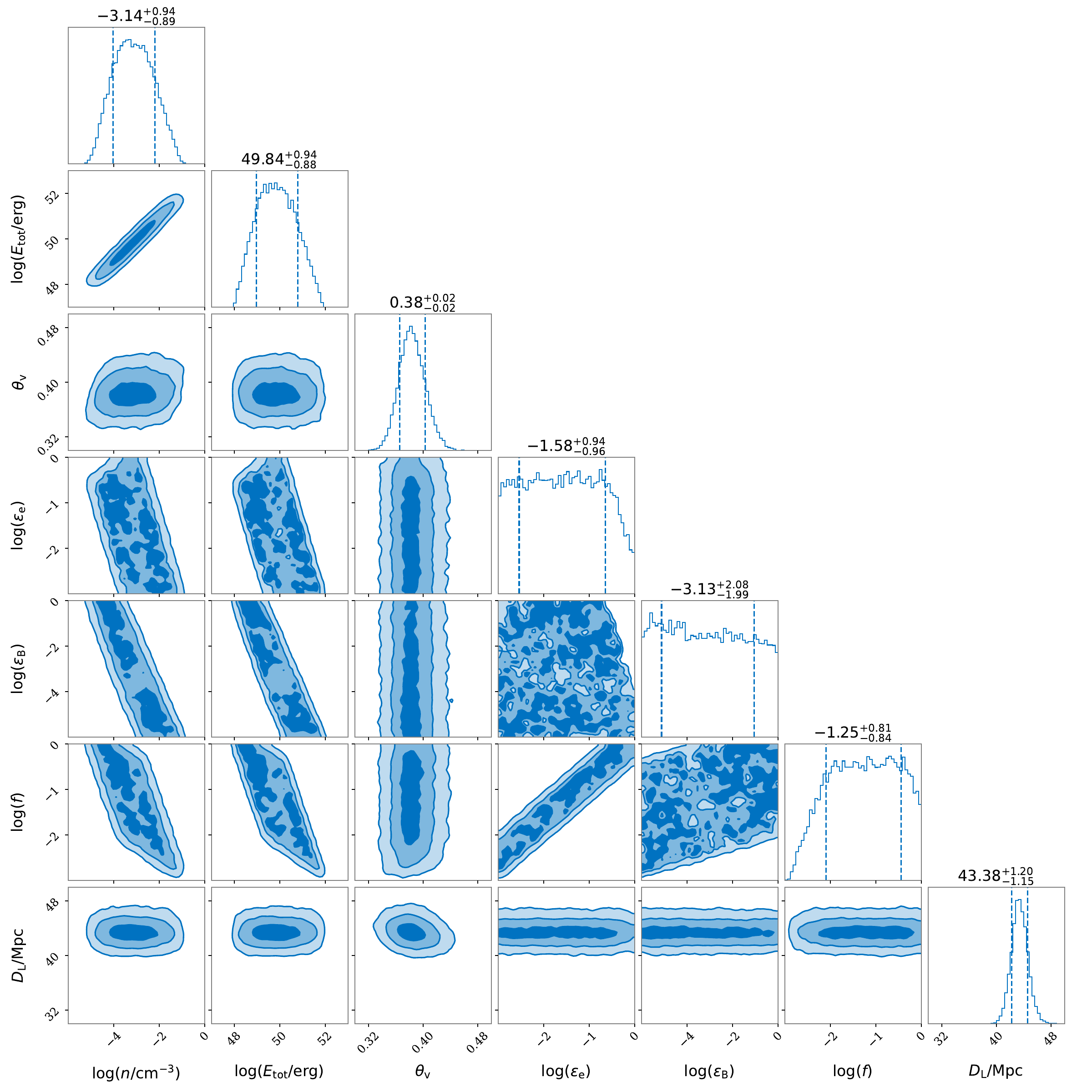}
    \caption{Parameter estimation result with prior set 3, where all parameters are left free. The prior of $\theta_{\rm v}$ and $D_{\rm L}$ follows Table \ref{tab:prior}. }
    \label{fig:pe_gw}
\end{figure}

\begin{figure}
	\includegraphics[width=\columnwidth]{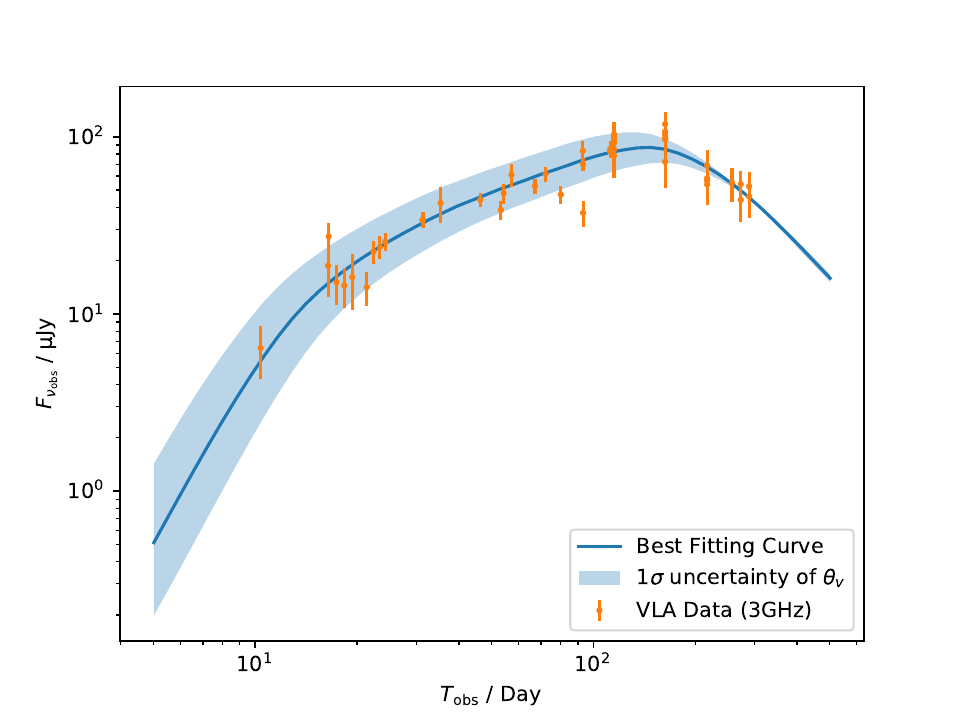}
    \caption{The light-curve fitting of the afterglow of GRB 170817A with $1\sigma$ uncertainty region of $\theta_{\rm v}$. All data points are normalized to $3\mathrm{GHz}$.}
    \label{fig:fitting}
\end{figure}

For each prior setup, we implement a nested sampling algorithm to generate posterior samples, employing the open package Pymultinest \citep{pymultinest}. We plot the parameter estimation results of different prior sets in Figure \ref{fig:pe_fixed_f}, \ref{fig:pe_free_f} and \ref{fig:pe_gw} respectively. In addition, we draw the best fitting light curve of prior set 3 together with radio data normalized to 3GHz in Figure \ref{fig:fitting}. 

\section{Implications}
\label{sec:results}

\subsection{Model Parameters}
The fitting process provides intriguing constraints to the parameters. All our fittings result in a tight constraint on the observing angle $\theta_{\rm v}=0.38\pm 0.02$ radian, which arises from the sensitive relation between the angle and light-curve slope. Our inferred value for $\theta_{\rm v}$ is consistent with the angle estimated by the jet's superluminal motion \citep{Mooley2018}. 

Since there are more model parameters than light-curve features, some parameters are degenerate \citep{2019ApJ...883...15G} as shown in Fig. \ref{fig:pe_fixed_f}-\ref{fig:pe_gw}. For example, it is not surprising that we cannot precisely constrain the magnetic field parameter $\epsilon_{\rm B}$, because the cooling frequency $\nu_{\rm c}$ is not observed in radio band. In fact, a further constraint on $\nu_c$ can be placed since the spectrum shows a single power-law function from the radio to X-ray band, which means the cooling break lies beyond the X-rays \citep{vla3}. The constraints of the afterglow parameters reported here (and the best fitting parameter in Figure \ref{fig:fitting}) are compatible with these studies.\footnote{Note that we did not include the Synchrotron Self-Compton effect in this work, which may have a modest effect to the electron cooling break \citep{2015MNRAS.454.1073B}.} The parameter $\nu_{\rm c}$ in our study therefore only plays a role of a scale factor in the calculation and does not affect our estimation of $\theta_{\rm v}$. However, it still helps us to extract some information of the electron acceleration process which is implied in Figure \ref{fig:pe_fixed_f} and \ref{fig:pe_free_f}. Assuming that $\epsilon_{\rm B}$ is not too high ($\epsilon_{\rm B}<0.01$, see the X-ray constraint from \cite{Hajela2019}), the $\epsilon_{\rm e} - \epsilon_{\rm B}$ contour plot in Figure \ref{fig:pe_fixed_f} shows a very large estimation of $\epsilon_{\rm e}$ ($>0.4$) which is on the high side from typically inferred gamma-ray burst afterglow parameters \citep{Kumar2015,2017MNRAS.472.3161B}. 

In fact, such a high estimation of $\epsilon_e$ results from the early-time observation when the frequency is possibly near or lower than $\nu_{\rm m}$ (e.g. The radio data of VLA at $\sim 16.5$ day shows a higher flux at higher frequency\citep{vla1} which implies a possibly positive spectral index). However, a relatively high $\nu_{\rm m}$ at very early time does not necessarily mean that a high amount of internal energy is transferred to non-thermal electrons. If we relax the assumption that all electrons in the shock region are accelerated (i.e., setting the fraction parameter $f$ as a free parameter), the fitting results in Figure \ref{fig:pe_free_f} show that a high estimation of $\epsilon_{\rm e}$ is no longer required, although the constraint is much loosened because of the strong degeneracy between $f$ and $\epsilon_{\rm e}$, as we can see from Eqation \ref{eq:gamma_m}. It can be regarded as an evidence showing that $f=1$ is disfavored. This parameter, however, does not affect the observing angle estimation.

The fitting after leaving $f$ free implies an outflow with initial kinetic energy of $E_{\rm tot}\approx 10^{49}-10^{51} \rm erg$ in a low density environment of $n\approx 10^{-4}-10^{-2} \rm cm^{-3}$. These results are compatible with the constraints placed on the ambient density by observations of the jet superluminal motion \citep{Mooley2018} and X-ray emission in the host galaxy \citep{Hajela2019}. 

\subsection{Breaking the $D_{\rm L}$-inclination degeneracy}

\begin{figure}
	\includegraphics[width=\columnwidth]{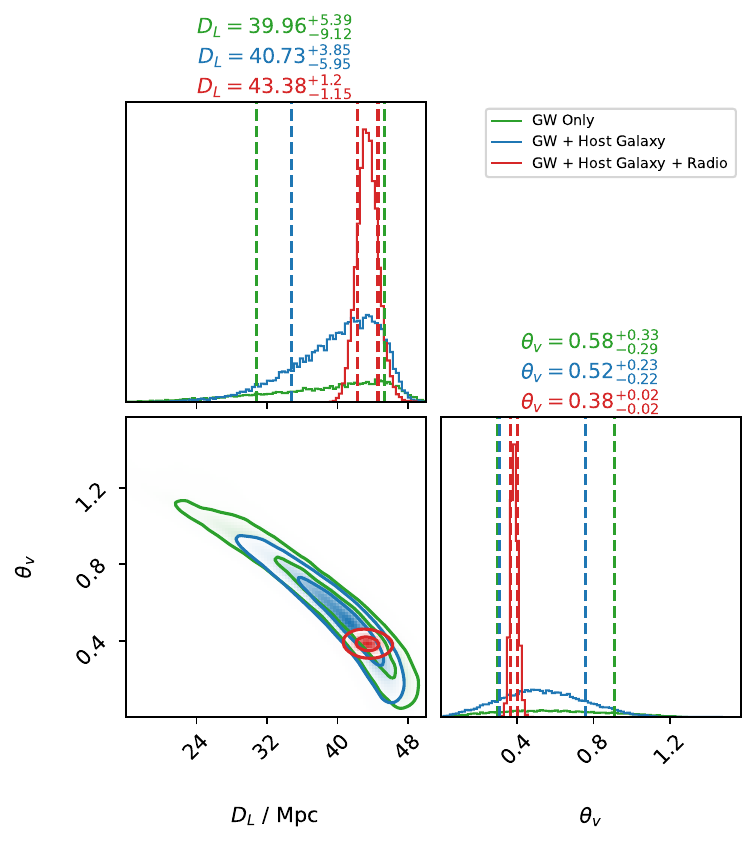}
    \caption{Contour plot of posterior probability distribution of $\theta_{\rm v}$ and $D_{\rm L}$ of GW170817. The green contour is from pure gravitational wave result. The blue contour is also from gravitational wave but with sky localization fixed to its host galaxy. The red contour is the joint fit using our model with radio afterglow observation and gravitational waves. It is the same contour as in Figure \ref{fig:pe_gw}. Inner and outer contour represents $1\sigma$ and $2\sigma$ uncertainty levels, respectively. Note that the EM data greatly reduce the uncertainty in both $\theta_{\rm v}$ and $D_{\rm L}$.}
    \label{fig:dl_theta}
\end{figure}

The tight constraint of $\theta_{\rm v}$ implies that, if we assume $\theta_{\rm v}$ being the same as the binary inclination (i.e., the jet axis is perpendicular to the merger plane), it can be used to break the degeneracy between luminosity distance and inclination angle from the gravitational wave parameter estimation. After applying prior set 3, we can see from Figure \ref{fig:pe_gw} that luminosity distance can be tightly constrained to $D_L=43.4\pm 1\ \rm Mpc$. We plot the results along with those from gravitational waves and host galaxy in Figure \ref{fig:dl_theta}. The contour plot of pure gravitational wave uses the published posterior samples\footnote{https://dcc.ligo.org/LIGO-P1800370/public}\citep{2019PhRvX...9c1040A}. Benefiting from a good estimation of $\theta_{\rm v}$, the uncertainty of $D_{\rm L}$ in our result has been reduced by a factor of 7 compared with pure gravitational wave estimation, and by a factor of 4 if host galaxy localization is added. This result shows that taking advantage of multimessenger parameter estimation, a simulated jet structure model with a small opening angle uncertainty can significantly increase the precision of distance measurement.

\subsection{Hubble Constant}
Our tight constraint of luminosity distance which is independent of the cosmological model can be used to infer cosmology parameters. In particular, at the nearby universe, it is possible to set a better constraint of the Hubble constant compared with the standard siren method using gravitational waves data alone. For the case of GW170817 which lies fairly close, the Hubble constant measurement is insensitive to other cosmological parameters such as $\Omega_{\rm m}$ and $\Omega_{\Lambda}$. It is also unnecessary to distinguish between different cosmological distance scales, since their difference is at the order of $v_{\rm H} / c \sim 1\%$ at such a distance, where $v_{\rm H}$ is the Hubble flow. We then follow the same method described in \cite{2017Natur.551...85A}, where the following expression provides a good approximation
\begin{equation}
    v_{\rm H} = H_0\cdot D.
\end{equation}

\begin{figure}
	\includegraphics[width=\columnwidth]{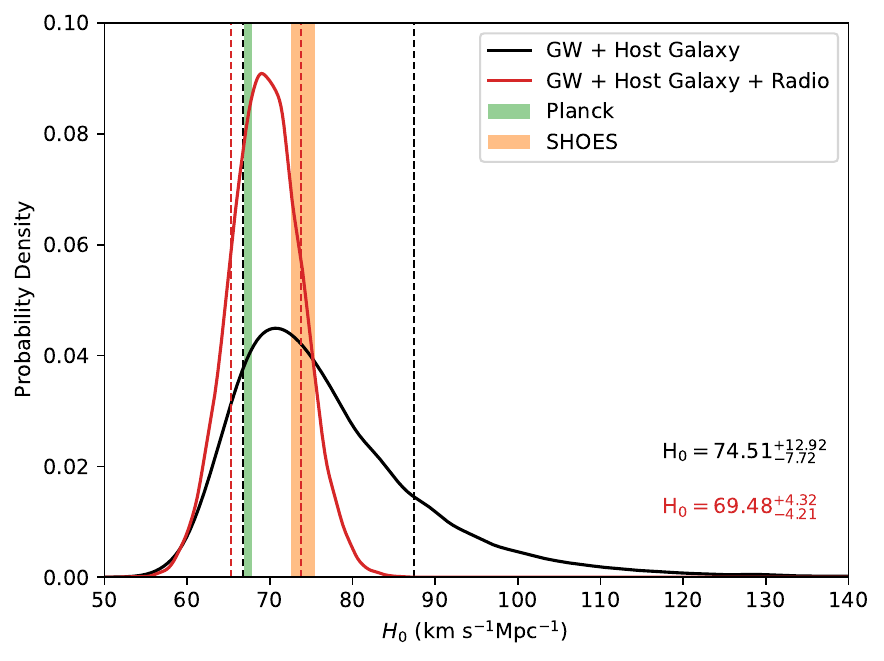}
    \caption{Hubble constant estimation using luminosity distance posteriors from our joint fitting. The green region represents the measurement from Planck Mission \citep{planck} and the orange one is from SHOES\citep{SHOES2019}. All uncertainty levels here are at $1\sigma$. The uncertainty in reduced by a factor of 2-3 by the inclusion of the radio afterglow constraints.}
    \label{fig:hubble}
\end{figure}

To make a comparison with standard siren method, we use the same Hubble flow $v_{\rm H} = 3017\pm 166\ \rm km\ s^{-1}$\citep{2007ApJ...655..790C} as in \cite{2017Natur.551...85A}, and assume it follows a normal distribution. The result is shown in Figure \ref{fig:hubble}, where we also plot the $1\sigma$ confidence interval of results from Planck Mission \citep{planck} and SHOES \citep{SHOES2019}. We constrain the Hubble constant to $H_0=69.5\pm 4\ \rm km\ s^{-1}Mpc^{-1}$ and the uncertainty has reduced by a factor of more than 2 by the inclusion of the radio constraints. It is worth mentioning that half of the uncertainty comes from the uncertainty in the peculiar velocity of the host galaxy in the Hubble flow which potentially could be constrained better in future observations or with methods incorporating such correction \citep{2019arXiv190908627M}. 

\section{Prospects of future detections}
\label{sec:prospect}
The prospects of multimessenger joint fitting of future gravitational wave events strongly depend on whether electromagnetic (EM) counterparts can be observed. In particular, in our model, the key factor is the detectability of the afterglow. Because of the extended jet structure, more events can be observed at larger angles \citep{2018ApJ...857..128J}. To quantify the prospects of afterglow detection of future binary neutron star merger events, we perform a Monte Carlo simulation that generates a million of candidate events uniformly distributed in the universe, and pick up those that can be detected by LIGO and Virgo. The method, which for simplicity focuses on the radio afterglow emission, is described below.

We generate samples of merger events produced by equal massive binary neutron stars $M=1.4M_\odot$ without spin. The samples are uniformly located in the volume up to a distance of 400 Mpc. Since the signal-to-noise-ratio(SNR) of a binary neutron star merger is dominated by it's inspiral stage, for simplicity, we adopt gravitational wave template TaylorF2 \citep{2011PhRvD..83h4051V} which provides good approximation at this stage. The detectors' sensitivity curve is adopted from the designed sensitivity of LIGO's public document\footnote{https://dcc.ligo.org/LIGO-P1200087-v47/public}, and we use it as an approximation to the O4 sensitivity. In the simulation an event is regarded as a confirmed detection if the LIGO and Virgo network SNR is larger than 12. We use open source package Bilby \citep{2019ApJS..241...27A} to perform signal injection and SNR calculation.

To test if those events can produce detectable radio emission, we use our model to generate light curves with parameters same to the best fitting curve in Figure \ref{fig:fitting}, except that distances and observing angles are free. Assuming that the outflow of all future events has the same intrinsic properties as GRB 170817A, such a setup can well estimate the flux of GRB 170817A-like events that are randomly distributed and oriented in the universe. The radio telescopes' detection limit is somewhat hard to estimate due to calibration issues. Considering that the first few radio data points of GRB 170817A detected by VLA are about $10-20\ \mu$Jy \citep{vla1}, we set two flux density limits: $10\rm \mu$Jy and $20\mu$Jy in our calculation. Another limitation to the radio telescopes is their response time, which depends on the progress of searching for the host galaxy. Here we set the initial time to be 1 day which is similar to the case of GW170817. Given the above setup, we can calculate the maximum distance for the detection of an event as a function of its observing angle. Here we assume that a successful detection happens when the peak flux density exceeds the telescope's detection limit.

\begin{figure}
	\includegraphics[width=\columnwidth]{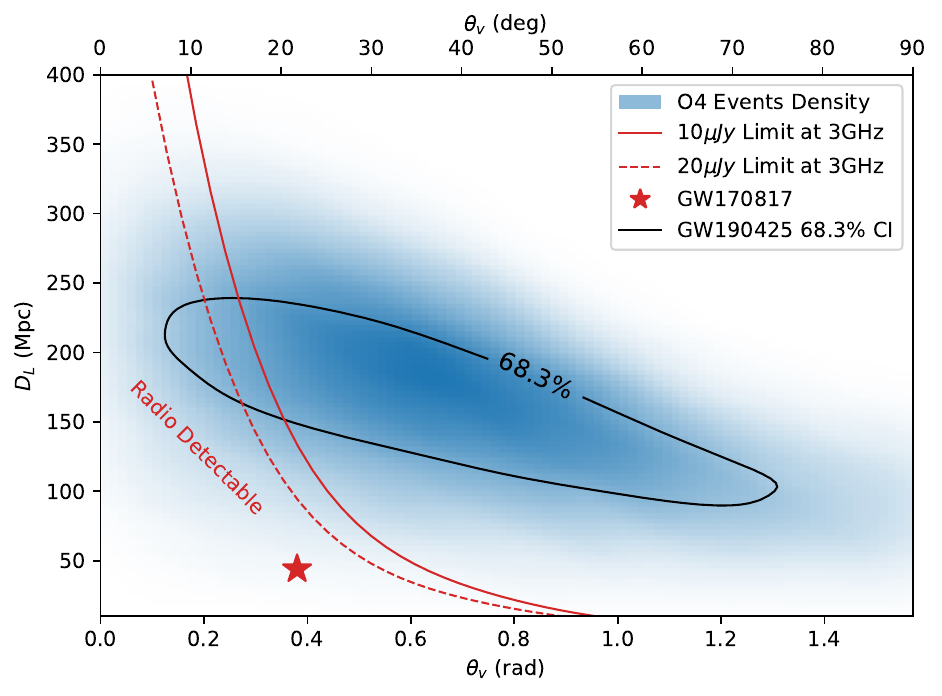}
    \caption{Prospects of O4 multimessenger detection. Here we assume the outflow of all future events having the same intrinsic properties as GW170817. The simulated gravitational wave events are represented by the blue density plot, where deeper color means high gathering event numbers. The red lines are maximum detection distance assuming $\rm 10\mu Jy$ or $\rm 20\mu Jy$ flux density limit of radio telescopes. Samples located on their low-left side can produce detectable radio afterglows in our model. }
    \label{fig:scatter}
\end{figure}

Our result is shown in Figure \ref{fig:scatter}. The simulated events' number density at each coordinate point are represented by the color depth. For those facing-off events (i.e. $\theta_{\rm v}>\pi/2$), we adjust the inclination angles to $\pi-\theta_{\rm v}$ to match the observing angle.

Our result shows that a considerable number of events can be detected at larger distance even beyond the average horizontal distance \citep{prospect} because of their optimal orientation. Our simulation estimates that detection rate of binary neutron star merger is approximately $R_{\rm GW}=20\left(\frac{R_{\rm BNS}}{1000\ \rm Gpc^{-3}yr^{-1}}\right)\mathrm{yr^{-1}}$, where $R_{\rm BNS}$ is the binary neutron star merger volumetric rate. Among those detections, according to different flux density limit setup, $10\% - 15\%$ may produce detectable radio counterparts at their peak emission. Using the latest estimation of $R_{\rm BNS}$ \citep{GW190425}, those numbers imply that we may have a few electromagnetic counterparts during LIGO O4. However, we should note that further upgrades of gravitational wave detectors' sensitivity may not help to substantially increase the number of electromagnetic counterparts, because the flux of faraway off-axis gamma-ray burst afterglows will drop below detection threshold of any current radio telescope. 

Radio emission is not the only EM counterpart expected in these sources. Other work has shown that the prompt emission of gamma-ray bursts lying in the radio detectable region are also likely detectable \citep{Kathir2018,lesson}, just as the case of GRB 170817A. Optical radiation from kilonovae whose radiation is more isotropic also helps from localization. The estimated fraction of events with EM counterparts is therefore robust. 

Considering the telescope response time caused by searching, we may still not be able to constrain the observing angle as well as GW170817 in events that are more distant. As we can see from Figure \ref{fig:scatter}, most events with EM counterparts have relatively low observing angle of $0.1-0.3$ rad or 5-15 deg and high luminosity distances of $100 - 250 \mathrm{Mpc}$. Those events, as we can see from Figure \ref{fig:lc}, may have declining afterglow emission, and are detectable only at very early time. The window of observing those events is limited so we may not be able to gather enough data points as we did for GW170817. GW170817, as we marked in the figure, is therefore a fairly unique event. 

However, we should expect that the observing angle constrained by EM data is still much better than using GW data only. To show this, we plot the parameter estimation contour plot of a newly detected event GW190425 here \citep{GW190425} which has no confirmed EM counterpart yet. We should then expect that for future detection, as long as its afterglow is detected, we can break the distance-observing angle degeneracy fairly well. 

\section{Conclusion and Discussion}
\label{sec:conclusion}
The blastwave driven by the gamma-ray burst jet is highly anisotropic. As a result, 
the light curves of gamma-ray burst afterglows strongly depend on the observing angle and can 
thus be used to constrain it. This constraint requires an accurate model for the 
jet structure. In this work, we use the structured jet model predicted  
by 3D GRMHD simulations of a black hole, torus system for the central engine of
neutron star mergers. We apply this model to gravitational and electromagnetic wave 
observations of GW170817, leading to a very 
tightly constrained observing angle for the system of $\theta_{\rm v} = 0.38\pm 0.02$ rad. Our model is essentially one-dimensional since we treat the shock evolution of different directions as being independent, and neglect the sideways expansion caused by pressure gradients along the shock surface. To include the sideways effect one needs a two-dimensional model. Such effects are important when the afterglow has peaked and the jet has sufficiently decelerated. 
The sideways expansion will result in a faster decrease of light curve after the peak. Therefore, the late-time afterglow can only be well 
described in models where such effects are included (see, e.g., \cite{2018MNRAS.481.2581L}). 
However, because in our analysis the observing angle is constrained by the light-curve features before the peak, the sideways jet spreading does not affect our result.

With a tightly constrained observing angle, we break the degeneracy of luminosity distance and 
inclination angle in gravitational waves parameter estimation, leading to a much better distance 
measurement $D_{\rm L}=43.4\pm 1$ Mpc. This result is independent of cosmology parameters and can be 
applied to measure the Hubble constant. In our work, we constrain the Hubble to be $H_0=69.5\pm 4\ \rm km\ s^{-1}Mpc^{-1}$. 

We also use Monte Carlo simulation to explore the prospects of applying this method to future 
detection of merger events. Assuming the current 
estimation of binary neutron star merger rate of $\sim 1000\ \rm Gpc^{-3}yr^{-1}$, there 
may be a few neutron star merger events per year with EM 
counterpart observed in LIGO's O4 stage (out of $\sim 20$ gravitational wave only events). 
The inclination angle of these events is likely to be in the range of 
0.1-0.3 rad and their distance will range from $\sim$100 Mpc to 300 Mpc. 
In a fair fraction of these events, our line of sight will be at the 
edge of the jet core, which is a different setup in comparison to the more inclined
GRB 170817A. This occurrence may increase the difficulty of accurately constraining 
observing angle, but the constraints will still be much better than using gravitational waves only. 
With several such events, the Hubble constant measurement in this method may achieve a higher 
precision and help to resolve the existing tension that appears among other methods.

Our results imply that the inclination angle of the merger can be very well constrained, provided that
we have a reliable model for the jet. The inference of the observing angle is most sensitive to the geometry, 
especially the opening angle of the jet. \cite{2019ApJ...883...15G} and \cite{2020arXiv200501754N} argued that the afterglow light curve before and around the peak sensitively relies on the ratio between observing angle and jet opening angle, which indicates another degeneracy. The estimated observing angle therefore relies on assumptions for the nature of the central engine. For the case of binary neutron star merger, considering the current constraint of neutron star maximum mass 
$M_{\rm NS, max}\approx 2.2M_{\odot}$\citep{2017ApJ...850L..19M, 2020PhRvD.101f3029S}, it is reasonable to expect that the merger remnant of GW170817 
(and of most binary neutron star mergers) collapses fast into a spinning black hole. Material with large amounts of 
specific angular momentum will surround the black hole forming a geometrically 
thick accretion disk or torus. The accretion of the torus can result in magnetic flux accumulation onto the black hole.
In a such system (i.e. a magnetized black hole with high spin), the Blanford-Znajek process seems 
to be responsible for launching a Poynting-flux dominated jet. The jet is then collimated by dense outflows from the 
disk until it breaks out from the surrounding gas. Here, we adopted the \cite{2019MNRAS.482.3373F} GRMHD setup in simulating 
this system. Although, clearly representing the state of the art, that work is limited to one model for the central engine.
Clearly, a large number of realistic simulations with different initial conditions, to account, e.g., for the different 
progenitor masses and mass ratios are needed to investigate whether the findings on the jet structure are robust.
Independent work \citep{lesson}, however, indicates a relatively narrow range of short gamma-ray burst opening angle, and its independence 
to the luminosity function. Together with the fact that our result is consistent with the constraint from superluminal motion, it supports the promise of our approach.

It is worth noting that the degeneracy between observing angle and jet opening angle can be broken by another constraint from the flux centroid motion observed in the radio band \citep{Mooley2018}. 
However, for future events, it might not be possible to break the degeneracy in such a way, since the expected event distances shown in Fig. \ref{fig:scatter} are far larger than GW170817, making the detection of flux centroid motion much harder. A Self-consistent 3D GRMHD simulations toward a reliable jet structure may then still be necessary.

\acknowledgments

We acknowledge support from the NASA ATP NNX17AG21G, the NSF AST-1910451 and the NSF AST-1816136 grants.

\bibliography{Afterglow}
\bibliographystyle{aasjournal}

\end{document}